\documentclass{aa}

\usepackage{txfonts,graphicx}

\usepackage{natbib,multirow} 
\bibpunct{(}{)}{;}{a}{}{,} 

\newcommand{\ylm}{Y$_{\ell}^m$}
\newcommand{\kms}{km s$^{-1}$}
\newcommand{\msun}{M$_{\odot}$}

\begin{document}

\title{Rotation and Convective Core Overshoot in $\theta$ Ophiuchi}

\author{C.C. Lovekin\thanks{Email: catherine.lovekin@obspm.fr}  \and M.-J. Goupil}
\institute{LESIA, Observatoire de Paris-Meudon, UMR8109, Meudon,  France.}
\date{Received / Accepted}

\abstract
{Recent work on several $\beta$ Cephei stars has succeeded in constraining both their interior rotation profile and their convective core overshoot.  In particular, a recent study focusing on $\theta$ Ophiuchi has shown that a convective core overshoot parameter of $\alpha_{ov}$ = 0.44 is required to model the observed pulsation frequencies, significantly higher than for other stars of this type.}
{We investigate the effects of rotation and overshoot in early type main sequence pulsators, such as $\beta$ Cephei stars, and attempt to use the low order pulsation frequencies to constrain these parameters.  This will be applied to a few test models and the $\beta$ Cephei star $\theta$ Ophiuchi.  }
{We use the 2D stellar evolution code ROTORC and the 2D linear adiabatic pulsation code NRO to calculate pulsation frequencies for 9.5 \msun\ models evolved to an age of 15.6 Myr.  We calculate low order p-modes ($\ell \leq 2$) for models with a range of rotation rates and convective core overshoot parameters.  These low order modes are the same range of modes observed in $\theta$ Ophiuchi.  }
{Using these models, we find that the convective core overshoot has a larger effect on the pulsation frequencies than the rotation, except in the most rapidly rotating models considered.  When the differences in radii are accounted for by scaling the frequencies by $\sqrt(GM/R(40^{\circ})^3)$, the effects of rotation diminish, but are not entirely accounted for.  Thus, this scaling emphasizes the differences produced by changing the convective core overshoot.  We find that increasing the convective core overshoot decreases the large separation, while producing a slight increase in the small separations.  We created a model frequency grid which spanned several rotation rates and convective core overshoot values.  We used this grid to define a modified $\chi^2$ statistic in order to determine the best fitting parameters from a set of observed frequencies.  Using this statistic, we are able to recover the rotation velocity and convective core overshoot for a few test models.  We have also performed a ``hare and hound" exercise to see how well 1D models can recover these parameters.  Finally, we discuss the case of the $\beta$ Cephei star $\theta$ Oph.  Using the observed frequencies and a fixed mass and metallicity, we find a lower overshoot than previously determined, with $\alpha_{ov}$ = 0.28 $\pm$ 0.05.  Our determination of the rotation rate agrees well with both previous work and observations, around 30 \kms.}
{}

\keywords{stars: early-type - Stars: oscillations - Stars:individual ($\theta$ Oph) - Stars: rotation -
Stars: interiors}

\maketitle

\titlerunning{Rotation and Overshoot in $\theta$ Oph}
\authorrunning{C.C. Lovekin \& M.-J. Goupil}

\section{Introduction}

Recently, great progress in the asteroseismology of $\beta$ Cephei stars has been made thanks to extensive observational campaigns, which have allowed constraints to be placed on the interior properties of some of these stars.  For example, \citet{aerts03} and \citet{aerts04v836} have compiled and analyzed 21 years of photometry for V836 Cen, identifying six frequencies and their degree and order.  Subsequent modeling has placed constraints on the mass, age, metallicity, convective core overshooting ($\alpha_{ov}$) and internal rotation profile for this star \citep{dupret04}.  They find strong evidence for $\alpha_{ov}$ = 0.1 $\pm$ 0.05 in the absence of rotational mixing.  Although the rotation rate for this star is quite slow, around 2 \kms, the observed frequencies can not be matched with a uniformly rotating model.  

A second $\beta$ Cephei star, $\nu$ Eridani, has also been studied extensively with both photometric and spectroscopic campaigns \citep{handler04,aerts04nueri}.  Nine modes were detected for this star, including the radial mode and two $\ell$ = 1 triplets \citep{deridder04}.  Modeling of $\nu$ Eridani has also shown that non-uniformly rotating models are required \citep{pamyatnykh04, ausseloos04}.  As for V836 Cen, this star cannot be uniformly rotating to match the observed frequencies, and some convective core overshooting may be required.  There is also some indication that the interior chemical composition is not homogeneous, with Fe overabundant in the driving zone.  

Several other $\beta$ Cephei stars have also been successfully modeled.  Using observations by the MOST satellite, \citet{aerts06} were able to place constraints on the physical parameters of $\delta$ Ceti, including constraining the convective core overshoot ($\alpha_{ov}$ = 0.2 $\pm$ 0.05).  Seismic modeling of $\beta$ CMa has constrained the core overshoot to $\alpha_{ov}$ = 0.2 $\pm$ 0.05, as well as placing constraints on the mass and age of the star \citep{mazumdar06}.  

Most recently, \citet{briquet07} have successfully modeled  the $\beta$ Cephei star $\theta$ Ophiuchi using rigid rotation.
This star has been the subject of both photometric and spectroscopic campaigns \citep{handler05,briquet05}, with 7 frequencies identified.  These frequencies are thought to be the radial fundamental, one $\ell$ = 1 triplet, and three components of an $\ell$ = 2 quintuplet based on both spectroscopic and photometric mode identification.   Recent independent modeling agrees with the $\ell$ identifications, although in some cases they assign a different $m$ to the modes \citep{dd09}.  Spectroscopic observations were also used to determine the metallicity of $\theta$ Ophiuchi, with a best value of Z = 0.0114 using the new Asplund mixture \citep{briquet07}.  Based on these observations, the best fitting model for $\theta$ Oph has been found to have a mass of about 8.2 \msun, X$_c$ = 0.38, T$_{eff}$ = 22 260, log L/L$_{\odot}$ = 3.85 and a rotational velocity of about 30 \kms\ \citep{briquet07}.  

The convective core overshoot determined for $\theta$ Ophiuchi is surprisingly large, $\alpha_{ov}$ = 0.44 $\pm$ 0.07.  This result is more than double that found for similar $\beta$ Cephei stars, which have $\alpha_{ov}$ around 0.1-0.2.  It is possible that the unusually high overshoot is a result of 1D modeling, which does not take into account rotational effects on the evolution.  Although $\theta$ Oph is relatively slowly rotating for a B star, its rotation velocity , $vsini$ $\sim$ 30 \kms, is significantly higher than the stars discussed above, which have rotation velocities around 2  \kms\ (V836 Cen) and 6 \kms\ ($\nu$ Eri).  Rotation and convective core overshoot may be complimentary effects, in which case including rotation could reduce the amount of convective core overshoot required to match the observed frequencies.  The $vsini$ of $\theta$ Oph is about 30 \kms and the star is thought to be viewed nearly equator on \citep{briquet05}.  Although this is not particularly rapid rotation for a star of this type, the rotation should produce some effect on the structure and frequencies.  We have chosen to model $\theta$ Oph using models which are uniformly rotating on the ZAMS.  In this work, we use 2D stellar evolution and linear adiabatic pulsation codes to determine the effects of rotation and overshoot on pulsation frequencies.  By including a 2D treatment of rotation, we investigate whether the overshoot of $\theta$ Oph could be reduced while still matching the observed frequencies.

This paper is organized as follows.  In section \ref{method}, we discuss the stellar evolution and pulsation calculations.  In section \ref{structure} we recall the effects of convective overshoot on the structure of the star, and in section \ref{overshoot} we discuss the effects of rotation and overshoot on the observed frequencies in these models.   Using the resulting variation we attempt to determine the rotation rate and overshoot using the observed frequencies in section \ref{idit}.  As asteroseismic modeling is more commonly done using 1D models, we perform a ``hare and hound" exercise to determine how different the results from the two methods can be.  This exercise is discussed in section \ref{HH}.  Finally, we constrain the rotation and overshoot of $\theta$ Oph in section \ref{thetaoph}.  Our results are summarized in section \ref{conclusions}.

\section{Numerical Method}
\label{method}

Our stellar models are calculated using the 2D stellar structure code {\tt ROTORC} \citep{bob90,bob95}.  This code takes the conservation equations for mass, energy, 3 components of momentum and the composition together with Poisson's equation and solves them implicitly on a two dimensional finite difference grid using the Henyey method \citep{henyey}.  We use OPAL opacities \citep{opal} and equation of state \citep{opaleos}.  Unlike standard 1D stellar evolution codes, {\tt ROTORC} uses the fractional surface equatorial radius and the colatitude as independent variables.  The surface equatorial radius is determined by requiring that the integral of the density over the volume of the model equal the total mass.  The surface radius in the other angular zones is calculated by assuming the surface is an equipotential.  As the models evolve, angular momentum is conserved locally.  Although these models are uniformly rotating on the ZAMS, they become slightly differentially rotating as they evolve.  This 2D modeling allows us to calculate the structural changes produced by rigid rotation without making assumptions about the shape {\it a priori}.

\begin{figure}
\resizebox{\hsize}{!}{\includegraphics{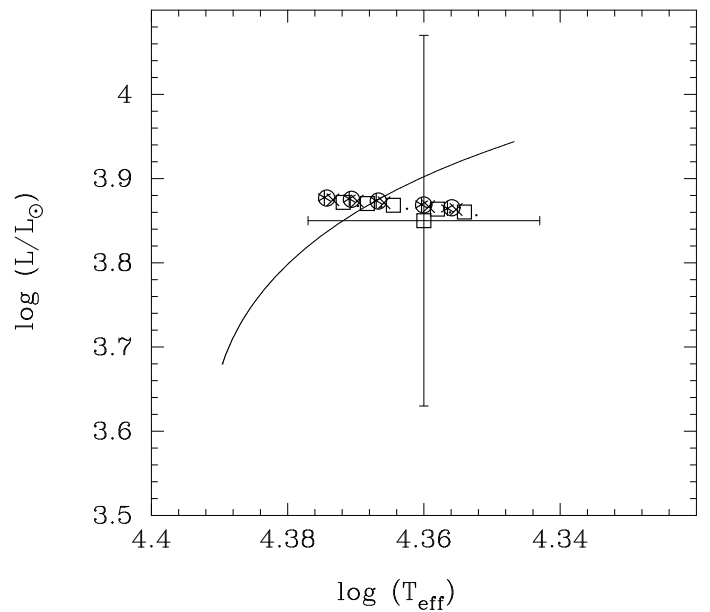}}

\caption{\label{fig:HR}  Location in the HR diagram for the 9.5 \msun\ models at age = 15.6 Myr, v$_{eq}$ = 0 (star), 50 (circle), 100 (X), 150 (square) and 200 (dot) \kms, $\alpha_{ov}$ = 0-0.38, increasing from right to left.  A sample evolution track is shown for a model with v$_{eq}$ = 0 and $\alpha_{ov}$ = 0.18.  The location and error bars for $\theta$ Oph are also shown, based on the photometric observations of \citet{handler05}.}
\end{figure}

Based on the calculations of \citet{briquet07}, the mass of $\theta$ Oph is around 8.5 \msun.  We have found that when rotation is taken into account, more massive models are generally needed to reach a given temperature and lumniosity.  We have calculated models at 8.5, 9 and 9.5 \msun, and have found that the 9.5 \msun\ models gave the best match to the observed temperature and luminosity of $\theta$ Oph.  We have calculated a small grid of these 9.5 \msun\ models evolved to an age of 15.6 Myr.  For these calculations, we used 581 radial zones and 10 angular zones.  The fractional surface equatorial radius was taken to be 1, and the ratio between the polar and equatorial radii decreases as the rotation rate increases.  For the velocities considered here, the rotational effects are not expected to be too severe \citep[see][]{me09}.  These models are uniformly rotating on the ZAMS, with surface equatorial velocities on the ZAMS of 0, 50, 100, 150 and 200 \kms, (corresponding to approximately 0, 35, 65, 110  and 150 \kms\ at 15.6 Myr) and overshoot parameters ($\alpha_{ov}$) of 0, 0.08, 0.18, 0.28 and 0.38.  We used a metallicity slightly higher than that determined by \citet{briquet07} for $\theta$ Oph, with Z = 0.02.  The initial hydrogen fraction of these models was 0.7.  The properties of these models are summarized in Table \ref{tab:models}.  The location of these models in the HR diagram is shown in Figure \ref{fig:HR}, along with a sample evolution track (v = 0, $\alpha_{ov}$ = 0.18).  The location of $\theta$ Oph with observational error bars from the photometric observations of \citet{handler05} is shown for reference. 

\begin{table*}
\caption{\label{tab:models}Summary of Model Properties}
\centering
\begin{tabular}{c c c c c c c c c}
\hline\hline
ZAMS v$_{eq}$ & Final v$_{eq}$ & $\alpha_{ov}$ & R$_{eq}$ & R(40$^{\circ}$) & R$_{p}$/R$_{eq}$ & X$_c$  & T$_{eff}$ & log(L/L$_{\odot}$)\\
(\kms) &  (\kms) &	& 	(R$_{\odot}$) & (R$_{\odot}$) & & & (K) & \\
\hline
0      & 0 &   0		&	6.757	& 6.757 &	1.0000	&  0.160 	&  21340	&  3.926	\\
0      & 0 &   0.08	& 	6.683	& 6.683 & 	1.0000	&  0.203 	&  21590	&  3.932	\\	
0      &  0 &  0.18	&	6.357	& 6.357 & 	1.0000	&  0.281 	&  22220	&  3.944	\\
0      & 0 &  0.28 	&	6.117	& 6.171 &	1.0000 	&  0.322 	&  22600	&  3.948	\\
0      & 0 & 0.38	&	6.004	& 6.004 &	1.0000	&  0.356  	&  22940	&  3.950	\\
\hline
50      &  35 &  0		&	6.790	& 6.784 & 	0.99850	&  0.158 	& 21300	& 3.927	\\
50      &  35 &  0.08	&	6.690	& 6.684 & 	0.99845 	&  0.203	& 21570	& 3.936	\\
50      &  36 &  0.18	&	6.363	& 6.357 & 	0.99836	&  0.281 	& 22210	& 3.943	\\
50      &  37 &  0.28	&	6.178 	& 6.172 & 	0.99832	&  0.322 	& 22590	& 3.747	\\
50      &  37 &  0.38	&	6.011	& 6.005 & 	0.99327	&  0.356 	& 22930	& 3.949	\\
\hline
100      &  69 &  0	&	6.793	& 6.769 & 	0.99389 	&  0.160 	& 21290 	& 3.925	\\
100      &  70 &  0.08	&	6.696	& 6.672 & 	0.99377	&  0.205 	& 21560	& 3.934	\\
100      &  72 &  0.18	&	6.375	& 6.350 & 	0.99352	&  0.282 	& 22180	& 3.941	\\
100      &  73 &  0.28	&	6.189	& 6.164 & 	0.99325	&  0.324 	& 22560	& 3.945	\\
100      &  74 &  0.38	&	6.024	& 6.000 & 	0.99312	&  0.357 	& 22890	& 3.947	\\
\hline
150      & 105 &   0	&	6.812	& 6.756 & 	0.98608 	&  0.163 	& 21250 	& 3.922	\\
150      & 106 &   0.08	&	6.713	& 6.657 & 	0.98582	&  0.208 	& 21520	& 3.931	\\
150      & 109 &   0.18	&	6.393	& 6.336 & 	0.98528	&  0.285 	& 22140	& 3.938	\\
150      & 110 &   0.28	&	6.211	& 6.155 & 	0.98473	&  0.326 	& 22510	& 3.941	\\
150      & 112 &  0.38	&	6.048	& 5.992 & 	0.98445	&  0.360 	& 22840	& 3.943	\\
\hline
200      & 141 &  0	&	6.853	& 6.750 & 	0.97504	&  0.165 	& 21190	& 3.919	\\
200      & 143 &   0.08	&	6.738	& 6.633 & 	0.97459	&  0.212 	& 21470	& 3.927	\\
200      &  146 &  0.18	&	6.425	& 6.321 & 	0.97318	&  0.288 	& 22070	& 3.933	\\
200      &  148 &  0.28	&	6.246	& 6.143 & 	0.97270	&  0.329 	& 22430	& 3.937	\\
200      &  150 &  0.38	&	6.088	& 5.984 &	0.97170	&  0.362 	& 22750	& 3.939	\\
\hline
\end{tabular}
\end{table*}

To calculate pulsation frequencies, we used a 2D linear adiabatic pulsation code, {\tt NRO} \citep{clement98}.  This code solves the linearized pulsation equations on a 2D grid using a finite difference technique.  The {\tt ROTORC} model, which is defined on a spherical polar grid is transformed into a model with the same number of radial zones, but defined on surfaces of constant density.  The pulsation equations are rewritten as finite difference expressions and the coefficients are placed in a band diagonal matrix.  Each element of this matrix is itself a matrix, containing the coefficients at each zone in the 2D grid.  The solution proceeds in two steps, from the centre outwards and from the surface inwards.  The solutions are required to match at some intermediate fitting surface.  At this point, a discriminant can be evaluated, which will only be satisfied (equal to zero) if an eigenvalue has been located.  Frequencies are detected by stepping through frequency space and looking for zero crossings in the value of the discriminant.  This method can result in missed frequencies if the frequencies are sufficiently close together, but this can usually be avoided by reducing the frequency step size.  For further discussion of the solution mechanism, refer to \citet{clement98,me08}.  

In rotating stars, a given mode cannot be described by a single spherical harmonic, but requires a linear combination of \ylm's.  We can calculate the first few terms in this combination as the eigenfunctions calculated with NRO are given as a function of $r$ and $\theta$, and are defined at several points on the surface.  Up to 9 angular zones can be included in the calculation, with one radial integration performed for each angle included.  The solution is given at N angles, which can subsequently be decomposed into the contributions of individual spherical harmonics, effectively calculating the first N terms in the linear combination.  Each radial integration contains angular derivatives, evaluated using finite differencing, so the resultant coupling among spherical harmonics arises naturally.  In {\tt NRO}, specifying $\ell$ specifies the parity of the mode, and the calculation is performed using the first N even or odd spherical harmonics.  In non-axisymmetric modes, the appropriate spherical harmonics are chosen starting with $\ell$ = $m$.

Since a given mode in a rotating star is described by a linear combination of spherical harmonics, the mode no longer has a unique $\ell$, although $m$ does remain unique.  Since $\ell$ is no longer unique, some new way of identifying modes must be used.  We have chosen to identify modes by the parameter $\ell_o$, which is the $\ell$ of the mode in the non-rotating model to which a given mode can be traced back.  This tracing is done for a series of models of gradually increasing velocity, based on both the shape of the latitudinal variation at the surface  and the frequency.  As the velocity gradually increases, the resulting distortion increases, allowing the modes to be identified.  However, this process becomes more difficult as the rotation rate increases, and is described in more detail in \citet{me08}.  Similar problems have been encountered by for example, \cite{reese09}.  

\begin{figure}
\resizebox{\hsize}{!}{\includegraphics{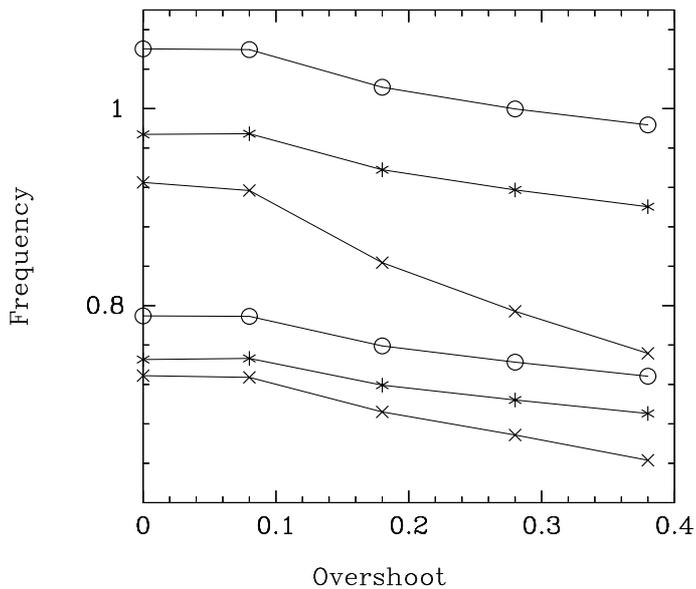}}

\caption{\label{fig:freqs}  The calculated frequencies for the non-rotating model; * - $\ell_o$ = 0, $\circ$ - $\ell_o$ = 1, x - $\ell_o$ = 2.  All frequencies are normalized by $\sqrt(GM/R(40^{\circ})^3)$, to account for the differences in stellar radius.  } 
\end{figure}

We have calculated frequencies for $\ell \leq 2$, as shown in Figure \ref{fig:freqs} for the non-rotating models.  We have chosen these low order p-modes for comparison with the modes detected in $\theta$ Oph.  The frequencies shown in this plot were scaled by a factor of $\sqrt(GM/R(40^{\circ})^3)$ to account for differences in the radii of the models.  We have chosen the radius at a colatitude of 40$^{\circ}$ as this radius has been found to be the radius most appropriate for calculating a pulsation constant in similar models \citep{me08}.

\section{Structural Effects}
\label{structure}

In order to better understand the influence of convective core overshooting, we have considered the effects of rotation and overshoot on the stellar structure for a collection of models at the same age.  As seen in Figure \ref{fig:HR} and Table \ref{tab:models}, increasing the convective core overshoot increases both the effective temperature and luminosity slightly.  Increased rotation does the opposite, causing a slight decrease in temperature and luminosity.  This decrease is small except for the most rapidly rotating model considered here (200 \kms).  These changes in temperature and luminosity can be related to the change in the radius of the star.  As the rotation rate increases, the equatorial radius increases, while the polar radius decreases.   This effect is well known, and has been demonstrated by many other authors \cite[see, for example][]{bodenheimer70,mm97,bob01}.  We have also found that for a given rotation rate, increasing the convective core overshoot decreases the radius by up to 0.8 R$_{\odot}$ for models at the same age and velocity, despite the corresponding increase in both temperature and luminosity (see Table \ref{tab:models}.)  

\begin{figure}
\resizebox{\hsize}{!}{\includegraphics{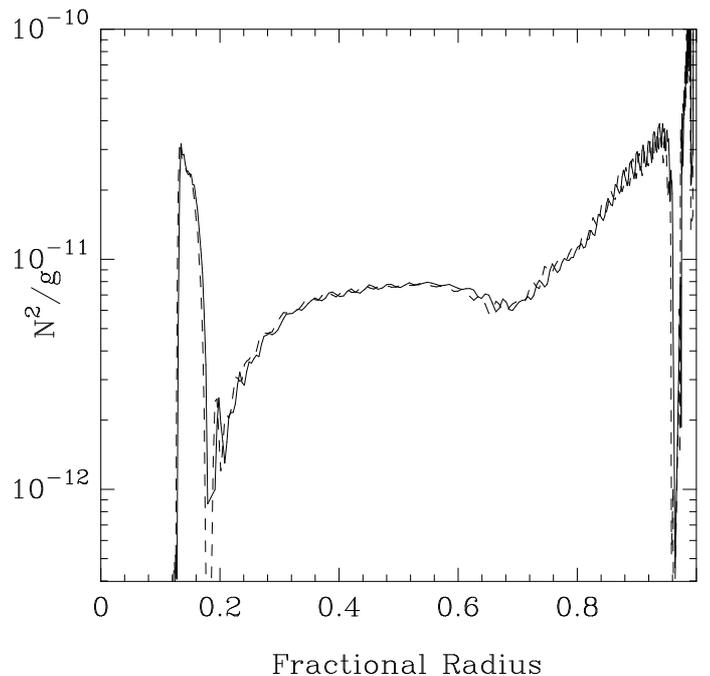}}

\caption{\label{bv1}The normalized Brunt-V\a{"}ais\a{"}al\a{"}a frequency for a 9.5 \msun\ non-rotating model (solid) and a model rotating uniformly at 200 \kms\ (dashed).  Both models have a convective core overshoot parameter of 0 and have been evolved to an age of 15.6 Myr.}
\end{figure}
\begin{figure}
\resizebox{\hsize}{!}{\includegraphics{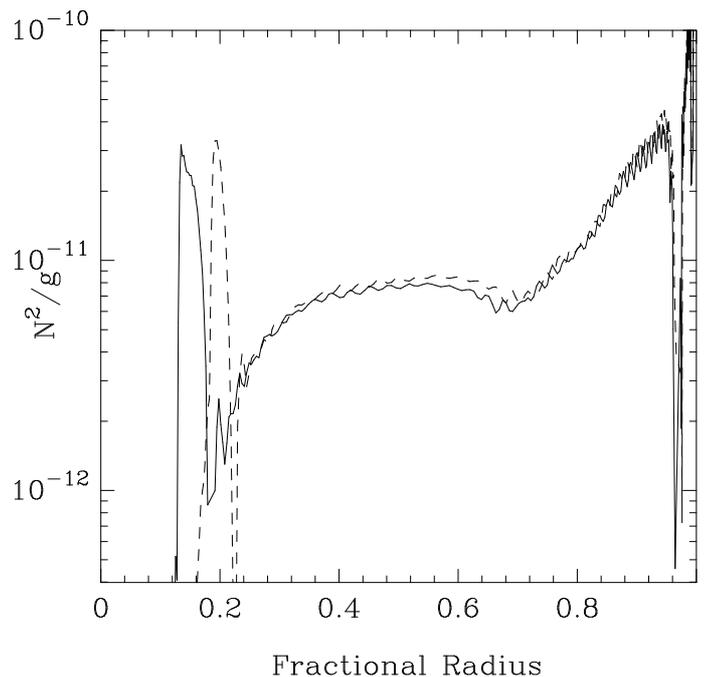}}
\caption{\label{bv2}The normalized Brunt-V\a{"}ais\a{"}al\a{"}a frequency for 9.5 \msun\ non-rotating models with core overshoot parameters of 0 (solid) and 0.38 (dashed), evolved to an age of 15.6Myr.}
\end{figure}

To help us assess the effect of changing core overshoot on the structure, and hence the frequencies, we have calculated the Brunt-V\a{"}ais\a{"}al\a{"}a frequency, defined as
\begin{equation}
N^2 = g\left[\frac{1}{P\Gamma_1}\frac{\partial P}{\partial r} - \frac{1}{\rho}\frac{\partial \rho}{\partial r}\right]
\end{equation}
for each of our models.  In Figure \ref{bv1}, we show $N^2/g$ for a non-rotating model and a rapidly rotating model (200 \kms) with no convective core overshooting.  The small wiggles visible, particularly in the outer sections of the star, are purely numerical effects, resulting from the finite difference calculation of the derivative.  The main peak at the boundary of the convective core has approximately the same size and shape in both models.  Clearly, changing the rotation rate does not produce much change in the shape of the normalized Brunt-V\a{"}ais\a{"}al\a{"}a frequency.  Although increased rotation increases the absolute size of the convective core, in terms of fractional radius, this size remains roughly constant.  In these models, local conservation of momentum means there is little radial mixing, so the composition gradient does not change significantly with rotation rate.  Although there are slight differences throughout the envelope of the star, they are small and unlikely to cause large shifts in the frequencies.  On the other hand, the overshoot has a much larger effect on the Brunt-V\a{"}ais\a{"}al\a{"}a frequency.  Figure \ref{bv2} shows the Brunt-V\a{"}ais\a{"}al\a{"}a frequency for two non-rotating models with core overshoot parameters of 0 and 0.38.  As expected, the peak at the boundary of the convective core has shifted outwards in radius.  This region is really the only significant difference, and the Brunt-V\a{"}ais\a{"}al\a{"}a frequencies are similar throughout the envelopes of the models.  
 
\section{Effects on the Frequencies and Frequency Separations}
\label{overshoot}

Using the 2D stellar evolution and pulsation calculations described in Section \ref{method}, we have calculated pulsation frequencies for models with rotational velocities from 0-200 \kms\ and overshoot parameters of 0-0.38.  Each model was evolved to an age of 15.6 Myr.  As the overshoot and rotation change the evolution slightly, all of these models have slightly different radii and core hydrogen fraction (X$_c$), as discussed in section \ref{structure} (see also Table \ref{tab:models} and Figure \ref{fig:HR}).  In order to offset the effects of this difference, we have scaled the frequencies by a factor of $\sqrt(GM/R(40^{\circ})^3)$ when looking for the effects of rotation and convective core overshoot.  We have chosen the radius at this colatitude as it has been found to produce the best pulsation constant \citep{me08}, and hence should be the best choice for scaling the frequencies.  

\subsection{Frequencies}

\begin{figure}
\resizebox{\hsize}{!}{\includegraphics{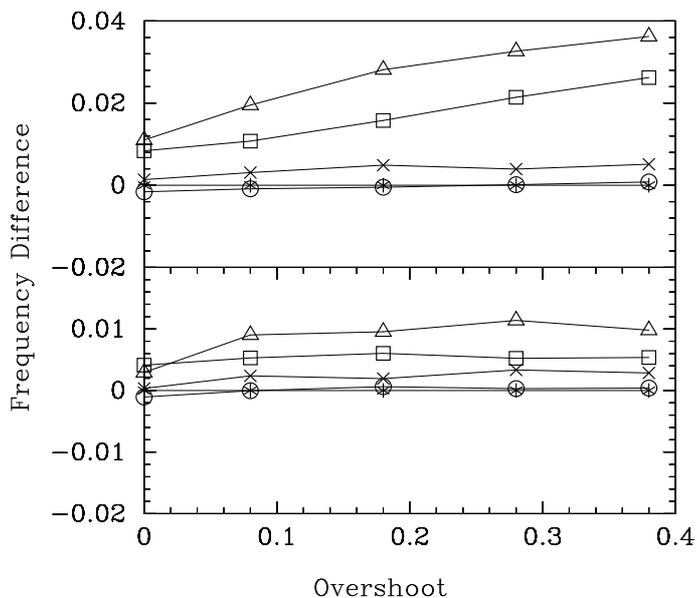}}

\caption{\label{fig:vdiff}The change in the $\ell_o$ = 0 normalized frequency relative to the frequency of the non-rotating model (stars) as a function of overshoot for models rotating at 50 \kms\ (circles), 100 \kms\ (X) 150 \kms (squares) and 200 \kms (triangles) for the p$_1$ (bottom) and p$_2$ (top) modes.}
\end{figure}

To determine the change in frequency produced by rotation and overshoot, we have plotted the scaled frequency as a function of overshoot relative to the non-rotating case (Figure \ref{fig:vdiff}) for the $\ell_o$ = 0 p$_1$ and p$_2$ modes.   The results are similar for the $\ell_o$ = 1 and 2 modes.  In all cases, the differences in scaled frequency produced by increasing the velocity are small, of order 1-1.5\%.   This is true regardless of which radius is used to scale the frequencies.  Using the equatorial radius does not change the results, while when the frequencies are scaled by the polar radius the magnitude of the frequency difference is slightly larger.  Interestingly, this change in scaling also changes the sign of the frequency differences.  When the frequencies are scaled by the equatorial radius, the frequencies increase as the rotation rate increases, while they decrease with increasing rotation rate when scaled by the polar radius.  

\begin{figure}
\resizebox{\hsize}{!}{\includegraphics{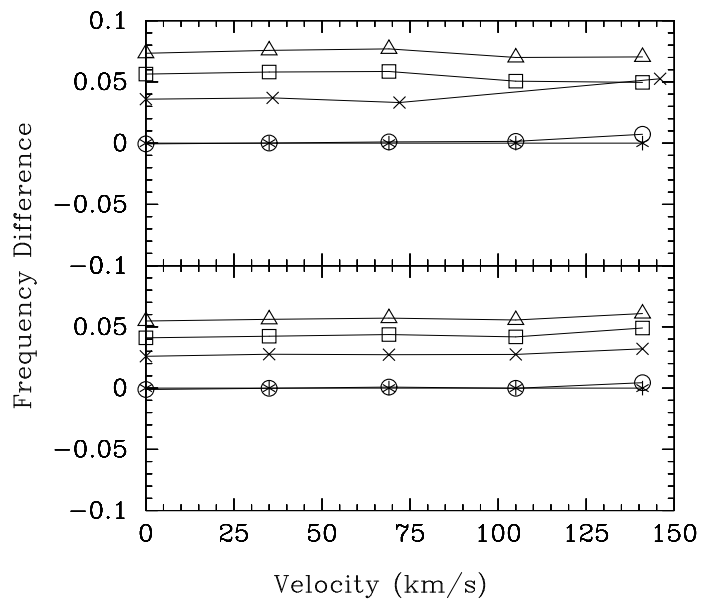}}

\caption{\label{fig:alphdiff}The change in the $\ell_o$ = 0 normalized frequency relative to the $\alpha_{ov}$ = 0 case (stars) as a function of velocity for the p$_1$ (bottom) and p$_2$ (top) modes.
Shown are overshoots of $\alpha_{ov}$ = 0.08 (circles), $\alpha_{ov}$ = 0.18 (X),  $\alpha_{ov}$ = 0.28 (squares) and $\alpha_{ov}$ = 0.38 (triangles).
}
\end{figure}

As discussed in section \ref{structure}, although the equatorial radius increases as the rotation rate increases, the polar radius can actually decrease slightly.  For example, while the equatorial radius in the models with no convective overshoot increases from 6.76 R$_{\odot}$ to 6.85 R$_{\odot}$ as the rotation rate increases from 0 to 200 \kms, the polar radius actually decreases to 6.68 R$_{\odot}$.  This increases the scaling factor as the rotation rate increases, and so the scaled frequencies actually decrease relative to the non-rotating case.  At a given overshoot, the radius at a colatitude of 40$^{\circ}$ stays nearly constant, changing by only 0.007 $R_{\odot}$ as the velocity increases from 0 to 200 \kms, giving us another reason to choose this as the scaling radius.  However, for a given velocity there is no point on the stellar surface at which the radius stays approximately constant for all overshoots, and scaling by the radius is not as effective.  In this way, our scaling accounts for the change in radius produced by rotation, but not that produced by convective core overshooting.  As a result, the curves shown in Figure \ref{fig:alphdiff} are relatively flat as a function of velocity, while the curves shown in Figure \ref{fig:vdiff} show variation with increasing overshoot.  This choice of scaling will highlight the differences resulting from the overshooting, making them easier to analyze.  

We have also calculated the change in frequency as a function of velocity relative to the case with $\alpha_{ov}$ = 0, shown in Figure \ref{fig:alphdiff}, again for the $\ell_o$ = 0 p$_1$ and p$_2$ modes.  The trends are similar, although larger in magnitude, with the differences typically of order 10 \%, as might be expected based on our choice of scaling.  In this case, overshoot appears to affect all $\ell_o$ values equally, as there is little variation with increasing $\ell_o$.  When scaling the frequencies, using the polar radius decreases the magnitude of the frequency difference slightly, but does not affect the sign of the frequency difference.  Changing the convective core overshoot does not affect the shape of the star as dramatically as does rotation, and both the polar and equatorial radii decrease as $\alpha_{ov}$ is increased.  
   
Both velocity and overshoot can significantly affect the absolute frequencies, although the effect is almost negligible (less than 1.5\%) with respect to rotation when the frequencies are scaled to account for differences in radius.  The differences shown in Figures \ref{fig:vdiff} and \ref{fig:alphdiff} are scaled using the radius at 40$^{\circ}$; the differences in the unscaled frequencies are about two orders of magnitude larger.  Although the scaled frequencies decrease with increasing rotation and overshoot, increasing overshoot actually increases the unscaled frequencies, while increasing velocity causes the frequencies to decrease.  Based on the results here, we find that the effects of slow to moderate uniform rotation (up to about 200 \kms) can be primarily accounted for by the effect on the stellar radius.  The effects of increasing convective core overshoot, on the other hand, are not so easily accounted for by scaling, and are nearly an order of magnitude larger than the rotational differences when scaled frequencies are considered.  Although the value of the frequencies themselves can change significantly, we find that as might be expected, overshoot produces no effect on the mode splitting.



\subsection{Frequency Separations}

\begin{figure}
\resizebox{\hsize}{!}{\includegraphics{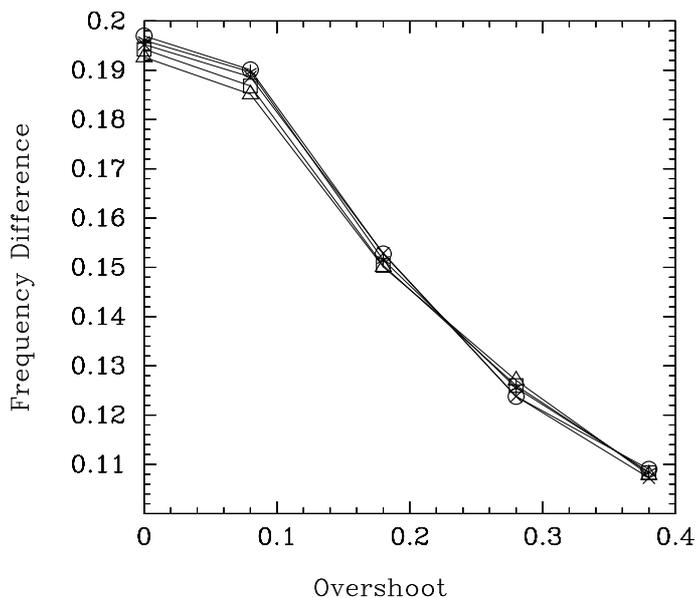}}

\caption{\label{fig:large}The large separation for the $\ell_o$ = 2 p$_1$ and p$_2$ modes as a function of overshoot for velocities of 0 \kms (stars), 50 \kms\ (circles), 100 \kms\ (X), 150 \kms\ (squares) and 200 \kms (triangles).  All frequencies are scaled by $\sqrt(GM/R^3)$ to account for differences in radius.  
}
\end{figure}

We have also calculated the effect of rotation and overshoot on the frequency separations.  Although the large and small separations are asymptotic limits, we have applied the definitions to our low order modes, and continue to refer to these as the large and small separation.  At large radial order, the large separation is defined as 
\begin{equation}
\Delta\nu_{\ell} = \nu_{\ell,n+1}-\nu_{\ell,n},
\end{equation}
where $n$ is the radial order of the modes.  This separation is shown in Figure \ref{fig:large} for the low order $\ell_o$ = 2 modes.  Both increasing rotation and increasing core overshoot produce a decrease in the large separation, most dramatically for the $\ell_o$ = 2 modes.  As the overshoot increases, the large separation decreases by about 7-10 \% for the $\ell_o$ = 0 and 1 modes, but falls by nearly 50 \% for the $\ell=2$ modes.   The decrease in large separation with increasing rotation rate, as found by \citet{me09} is also seen, although the effects are not as dramatic as the effects of convective core overshoot.  Again, this is at least in part due to our choice of scaling, which minimizes the effects of rotation to better show the effects of core overshoot.  

\begin{figure}
\resizebox{\hsize}{!}{\includegraphics{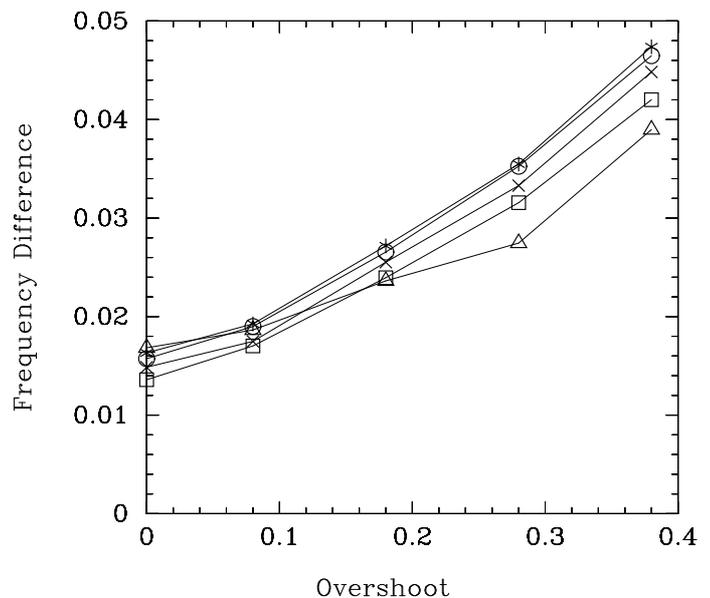}}

\caption{\label{fig:small}The small separations for even ($\ell_o$ = 0, 2, p$_3$ and p$_4$) modes as a function of overshoot for 0 \kms\ (stars), 50 \kms\ (circles), 100 \kms\ (X), 150 \kms\ (squares) and 200 \kms (triangles).  As for the large separation, the frequencies are scaled by $\sqrt(GM/R^3)$ to account for the different radii of the models.}
\end{figure}

The small separation is defined as 
\begin{equation}
d_{\ell,n} = \nu_{\ell,n}-\nu_{\ell+2,n-1}
\end{equation}
and shown in Figure \ref{fig:small} for the even modes.  As with the large separation, the small separation decreases slightly as a function of rotation rate up to approximately 200 \kms.  At this velocity, there is a minimum in the small separation, and the small separation increases by approximately 20 $\mu$Hz as the rotation increases beyond this velocity \citep{me09}.  The effects of overshoot on the small separation are shown in Figure \ref{fig:small}, were a significant increase can be seen as the core overshoot increases.  A large increase in the small separation could therefore be indicative of either a large core overshoot or a high rotation rate.   


\section{Constraining $\alpha_{ov}$}
\label{idit}

Once we had established the effects of rotation and overshooting on the pulsation frequencies and separations, we attempted to determine if these could be used to constrain a star given a set of observed frequencies.  To do this, we calculated three test models with the same mass and age as the previous grid  of models (see Section \ref{method}), but with slightly different rotation rates and convective core overshoots: (v$_{eq}$,$\alpha_{ov}$) = (80 \kms, 0.1), (180 \kms, 0.1) and (80 \kms, 0.3).  By comparing the differences between the frequencies for these test models (henceforth the observed frequencies) and the frequencies in our original grid (model frequencies) we hope to constrain the rotation and overshoot.  Note that throughout this section, we compare absolute frequencies, not scaled frequencies as in the previous section.  

\begin{figure}
\resizebox{\hsize}{!}{\includegraphics{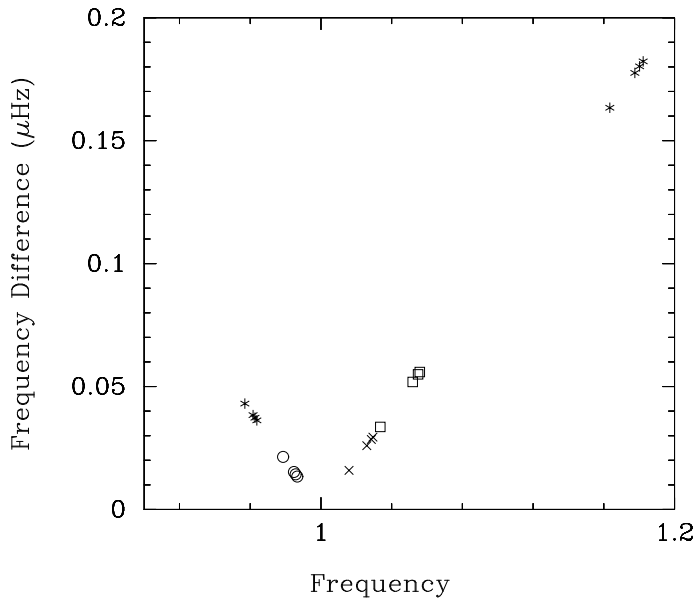}}
\caption{\label{fig:fit80}Determination of overshoot using the $\ell_o$ = 0 p$_1$ frequency for an observed model with $\alpha$ = 0.1, v = 80 \kms.  The observed frequency is compared to the corresponding frequency for each model in the grid, plotted against the observed frequency.  All frequencies are normalized by the observed $\ell_o$ = 0 frequency.  The models corresponding to constant $\alpha_{ov}$ cluster together in this diagram:  $\alpha_{ov}$ = 0 (stars), $\alpha_{ov}$ = 0.08 (circles), $\alpha_{ov}$ = 0.18 (X), $\alpha_{ov}$ = 0.28 (squares) and $\alpha_{ov}$ =0.38 (stars).  The best fit model has $\alpha_{ov}$ between 0.08 and 0.18.}
\end{figure}

Initially, we attempted to match individual frequencies.  For a given observed frequency, we calculate the absolute value of the difference between the observed frequency and each model frequency ($|\nu_{\ell,grid}/\nu_{\ell = 0,obs}-\nu_{\ell,obs}/\nu_{\ell=0,obs}|$) and plot this difference versus the observed frequency ($\nu_{\ell,obs}/\nu_{\ell=0,obs}$).  All frequencies are normalized by the observed $\ell_o$ = 0 frequency to offset the effects of varying radius.  The results of this calculation are shown in Figure \ref{fig:fit80} for the $\ell_o$ = 0 p$_1$ frequency of the observed v$_{eq}$ = 80 \kms, $\alpha$ = 0.1 case.  The models with the same convective core overshoot parameter cluster in this diagram, giving no information about the velocity.  There is a clear minimum difference in frequency which falls between $\alpha_{ov}$ = 0.08 and 0.18, corresponding to the target overshoot of 0.1.  

\begin{figure}

\resizebox{\hsize}{!}{\includegraphics{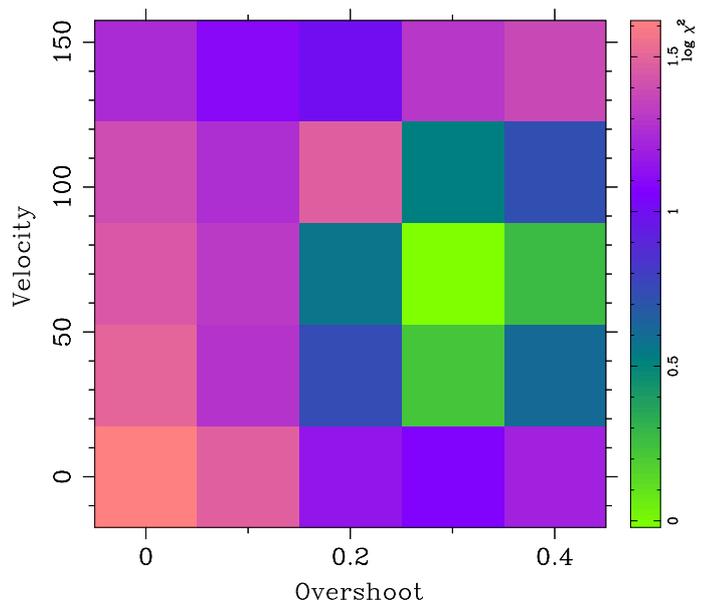}}
\caption{\label{fig:chi2}The log of $\chi^2$ for an ``observed" model with $\alpha_{ov}$ =  0.3 and $v_{eq}$ = 59 \kms.  The minimum can be seen around (65,0.28), with models at higher velocity and similar overshoot also acceptable.  Although the resolution is course, we can place some constraints on a given model.  }
\end{figure}

Using this method, we can of course find a good match for a single frequency.  However, we found that fitting different observed frequencies from a single observed star can give different results for the best fitting overshoot.  To determine the single best solution for the rotation and overshoot for each set of observations, we combine the data for all observed frequencies to determine the best  overall fit for a given model.  We do this using a modified $\chi^2$ statistic.   To calculate this, we compare the observed frequencies to the calculated frequencies from each model.  For a given rotation and overshoot, we determine the best fitting model frequency for each observed frequency, where the $\ell_o$ of the mode is known, but $m$ and $n$ are free parameters.  The total $\chi^2$ for this model is the sum of the differences squared for the best-fitting matches for the individual frequencies, normalized by the number of frequencies we fit:  
\begin{equation}
\chi^2 = \frac{1}{N}\sum_i(\nu_{the,i} - \nu_{obs,i})^2
\end{equation}
where N is the number of frequencies fit for each model.  The surface equatorial velocity and core overshoot of the best fit model is given by the lowest overall $\chi^2$.  The $\chi^2$ calculated for each model can then be plotted as a function of velocity and overshoot, as shown in Figure \ref{fig:chi2}, giving us a simple visual way of determining the best fitting model.  

This method works best if some additional constraints are applied.   Once a model frequency has been matched to an observed frequency, it is flagged and considered unavailable for future matches.  If more than one observed frequency is best matched to the same model frequency, we use a recursive algorithm to determine which arrangement of the frequencies gives the lowest $\chi^2$.  This also allows us to determine the best fit in cases where we have more observed frequencies than model frequencies.  We also exclude observed frequencies that are more than 5 $\mu$Hz above (below) the maximum (minimum) value in the model grid.  Both of these constraints remove frequencies from the set of observed frequencies, and different numbers of frequencies may be removed when comparing to different models.  To account for this, for each overshoot and velocity, we normalize the $\chi^2$ by the number of observed frequencies for which matches were found.  

Using this method, we were able to recover the input parameters for our three ``observed" models.  The observed models have velocities of 80 and 180 \kms\ on the ZAMS, which correspond to 57 or 65 \kms\ (depending on the core overshooting) and 129 \kms at an age of 15.6 Myr, and overshoot parameters of 0.1 and 0.3 H$_p$.  For a model with (v = 129 \kms, $\alpha_{ov}$ = 0.1H$_p$), we recovered (110, 0.08), with an uncertainty of $\pm$ 50 \kms\ in velocity and $\pm$ 0.1 in overshoot due to the coarseness of our model grid.  Similarly, for a model with (57, 0.1) we recovered (65, 0.18) and for a model with (59, 0.3), we recovered (65, 0.28).   A sample plot of the $\chi^2$ space is shown in Figure \ref{fig:chi2} for a model with parameters (59, 0.3).  In general, we find acceptable solutions over a range of bins, generally straddling the true value.  In the example shown in Figure \ref{fig:chi2}, the best solution has a slightly higher velocity and lower overshoot than the true values, and models with a slightly lower velocity or higher overshoot are also acceptable solutions.  

\begin{figure}

\resizebox{\hsize}{!}{\includegraphics{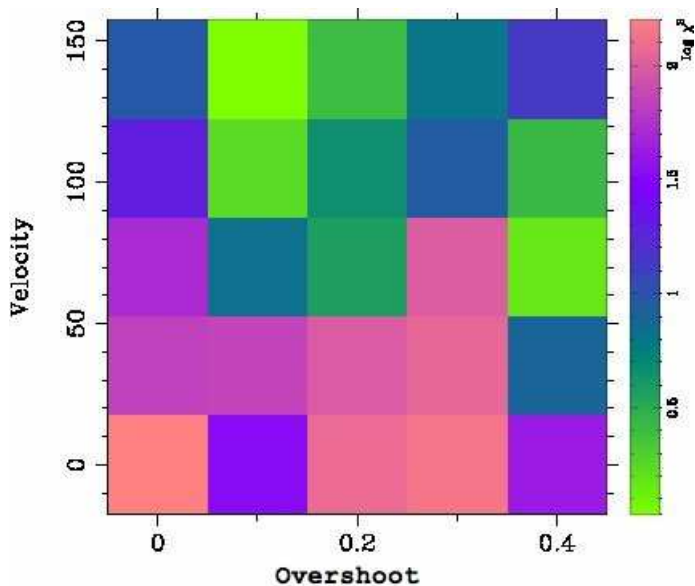}}
\caption{\label{fig:chishort}The log of $\chi^2$ for an ``observed" model with $\alpha_{ov}$ =  0.1 and $v_{eq}$ = 129 \kms, fit using 10 frequencies.  The best fitting solution can be seen at (145 \kms, 0.08), while a secondary minimum has appeared at lower velocity and higher overshoot.}
\end{figure}

The models shown above were calculated using a full set of calculated frequencies (about 20 frequencies), including the first two radial orders for all modes.  In more realistic cases, fewer frequencies would be available, and the $m$ and $n$ of these modes would be unknown.  To examine a more realistic case, we have created an ``observed" star with a reduced set of frequencies and compared them to the full model grid.  We have assumed the frequencies of the $\ell = 0$ fundamental, a $\ell = 1$ triplet and a full $\ell = 2$ quintuplet are known (10 frequencies in total).  We found that reducing the number of frequencies in this way did not affect the determination of the best fitting solution.  We do find some evidence for a degeneracy between rotation and overshoot in the most rapidly rotating case, shown in Figure \ref{fig:chishort}.  In this case, when only 10 frequencies are included, the best fitting solution remains in the same area, now at 145 \kms, $\alpha_{ov}$ = 0.08.  A secondary minimum has also appeared, with acceptable solutions having a core overshoot parameter of 0.38 and a velocity of 65 \kms.  

Clearly, our resolution, particularly in velocity, is very low, and better constraints could be established by using a finer grid.  However, even with this coarse resolution, the method successfully returns the correct (albeit uncertain) surface equatorial velocity and core overshoot, at least in cases where the mass and age of the star are known.  In general, the mass and age are not known, which will make the fitting process more difficult.

\section{Hare and Hound Exercise}
\label{HH}

Next, we decided to see how well this determination of rotation and overshoot could be done using 1D models.  This will help quantify the errors introduced by neglecting rapid rotation in these models.  To do this, low order frequencies were calculated by one of us (CCL) using a uniformly  rotating 2D model as described above.  The selected model is an 8.5 \msun model, evolved to an age of 20 Myr, with a ZAMS rotation rate of 150 \kms\ (108 \kms\ after 20 Myr).  The calculated frequencies and observed position in the HR diagram were passed on as a set of observational quantities, given in Table \ref{tab:hhobs}.  One of us (MJG) then used these observations to try and determine the convective core overshoot and rotation rate of the star.  The star was assumed to be a uniformly rotating $\beta$ Cephei star oscillating with low order p and g modes with metallicity Z = 0.02.  For comparison, we also use the frequencies in Table \ref{tab:hhobs} in a 2D fitting process as described above, assuming the mass and age of the star are known.

\begin{table*}
\caption{\label{tab:hhobs}``Observational" Data Used in Hare \& Hound Exercise}
\centering
\begin{tabular}{c c c c c c c }
\hline
log L/L$_{\odot}$ & T$_{eff}$  (K) &  & & & & \\
\hline
3.7 $\pm$ 0.1 & 21700 $\pm$ 150  & &  &   &  \\
\hline\hline
\multicolumn{2}{c}{$\ell $} &  $m=-2$ & $m=-1$ & $m=0$ & $m=1$ & $m=2$ \\ \hline
\multirow{2}{*}{$\ell = 0$} & p$_1$ & \ldots& \ldots& 86.78 &  \ldots&  \ldots\\
 & p$_2$ & \ldots& \ldots&113.05  & \ldots&  \ldots\\
\hline
\multirow{3}{*}{$\ell=1$} & p$_1$ &\ldots &75.054 & 68.367  & 64.936 &   \ldots\\
&p$_2$ & \ldots&95.205 & 91.550  & 85.952 &   \ldots\\
&g$_1$ & \ldots&38.577 & 35.318  & 29.693 &   \ldots\\ 
\hline
\multirow{3}{*}{$\ell=2$} & p$_1$ & 75.071 & 79.820 & 84.348 & 88.829 & 92.796 \\
& p$_2$ & 92.195 & 98.321 & 104.033 & 108.771 & 112.744 \\
& g$_1$ & 43.115 & 49.130 & 54.846 & 60.191 & 65.267 \\
\hline
\end{tabular}
\end{table*}

\begin{figure}
\resizebox{\hsize}{!}{\includegraphics{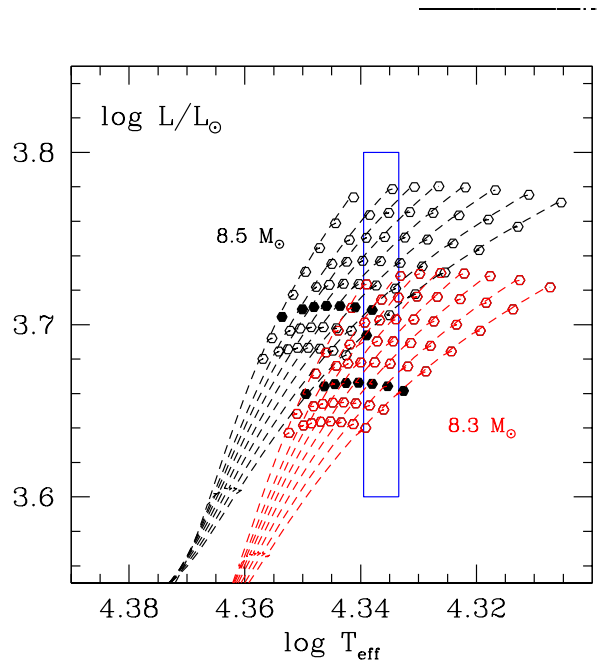}}
\caption{\label{fig:hareHR}HR diagram showing the evolutionary tracks (dashed lines) for 8.5 \msun\ (black) and 8.3 \msun\ (red) models with different values of the core overshoot parameter $\alpha_{ov}$.  Open dots represent models at time intervals of 1 Myr around an age of 20 Myr (filled dots).  The box represents the uncertainty of the observed location of the ``star'' in this diagram, as given in Table \ref{tab:hhobs}.}
\end{figure}

In the 1D case, this star was modelled with masses of 8.3 and 8.5 \msun, chosen based on which tracks cross the observed location in the HR diagram.  The calculated evolutionary sequences cross the error box at an age between 18 and 25 Myr, depending on the amount of convective core overshoot.  Sequences were considered for $\alpha$ = 0, 0.05, 0.1, 0.2, 0.3 and 0.4.  As expected, there is a degeneracy between mass, age and overshoot in the HR diagram, shown in Figure \ref{fig:hareHR}.   
The oscillation code  used in this section, WAR(saw)M(eudon),  includes the effects of uniform rotation as a perturbation  for axisymmetric modes as well as  frequencies of nonaxisymmetric modes up to third order in the rotation rate \citep{soufi98,ddetal02,goupiltalon09,goupil09,goupil10}.  
Comparison between frequencies obtained with a perturbed approach and a nonperturbative one  can be found in \citet{reese06,lignieres} for a frequency comparison using the same model, a 2D polytrope, in both approaches, and in \citet{rhita09} for a comparison using  a 1D polytrope in the perturbative approach
and a 2D polytrope involved in the nonperturbative one. 

\begin{figure}
\resizebox{\hsize}{!}{\includegraphics{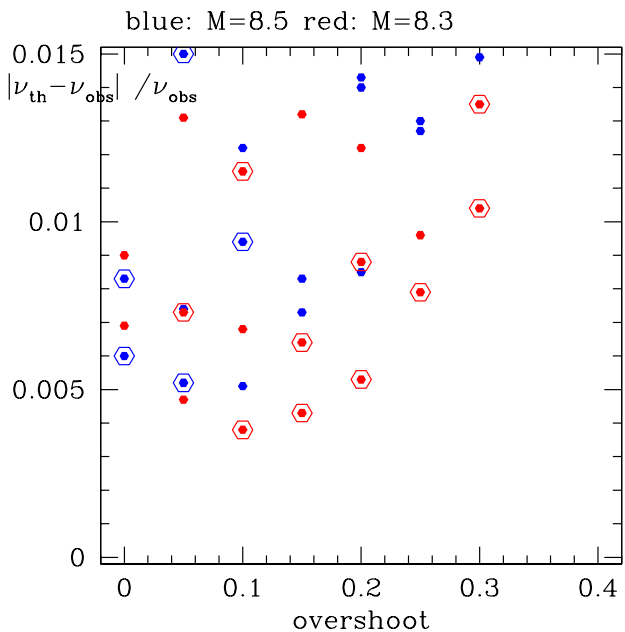}}
\caption{\label{fig:harechi} The quantity $\chi^2$ for modes ($\ell = 0$ $p_1$, $p_2$, and $\ell = 1$ $g_1$, $g_2$, and $p_1$) as a function of overshoot for models with mass of 8.5 \msun\ (blue) and 8.3 \msun\ (red). } 
\end{figure}

 Initially, frequencies were calculated assuming no rotation.
Particularly for the $\ell_o$ = 1 modes, matching the frequencies was a problem, and for some modes no solution was found.  In this case, the modes identified in Table \ref{tab:hhobs} as $p_1$, $p_2$ and $g_1$ had frequencies corresponding to the $p_1$, $g_1$ and $g_2$ modes in the 1D models.  None of the models considered were able to simultaneously fit all of the modes at the same age.  A $\chi^2$ technique, similar to that described above, was then used to fit all of the modes simultaneously.   This $\chi^2$ fit determined that the 8.3 \msun\ models gave a better fit to the data, with overshoots of around 0.1 $\pm$ 0.05.  The models with the lowest $\chi^2$ have ages between 21-22 Myr, and the temperature and luminosity are within the observed errors of the target ``star" (see Figure \ref{fig:harechi}).  

The 1D models used in this section are capable of including the effects of uniform rotation as a perturbation, allowing the mode splitting to be included.  A second $\chi^2$ was calculated for the splitting of the models compared to the ``observed" model.  As expected, this splitting was found to be independent of the overshoot.  Good agreement was found for rotation rates below 125 \kms, while above this rate the g mode splitting does not agree and $\chi^2$ becomes large.  

When all effects are included, the 1D modeling finds the best solution to be 8.3 \msun, 21 Myr, $\alpha_{ov}$ = 0.1 H$_p$ and v$_{eq}$ = 125 \kms.  In contrast, the 2D modelling described above finds a best fit model with $\alpha_{ov}$ = 0.28 and v$_{eq}$ = 145 \kms, given an assumed mass and age of 8.5 \msun\ and 20 Myr.  Solutions at 65 \kms\ are also acceptable, as are core overshoot parameters of 0.18.  The large range in possible velocities found with the 2D models are a result of the coarseness of the grid, which contains no models with velocities near 100 \kms.  Although the 1D models are able to determine the velocity of a target model, the determination of the overshoot is much less accurate.  As discussed in \citet{lignieres} and \citet{me08}, for stars rotating more rapidly than about 100 \kms, a more accurate treatment of the rotation and pulsation is needed.  These works have shown that a 2D approach tends to produce a larger effect on the frequencies than 1D perturbative methods.  As discussed above, rotation and overshoot change the unscaled frequencies in opposite directions.  While increasing rotation decreases the frequencies, increasing core overshoot increases the frequencies.  According to \citet{lignieres, me08}, 1D calculations will not find as large a shift from rotation as 2D models, and hence will need a smaller overshoot to match the observed frequency in rapidly rotating stars.    

\section{Application to $\theta$ Oph}
\label{thetaoph}

Finally, we decided to test the technique discussed in Section \ref{idit} on $\theta$ Oph.  As discussed in Section \ref{idit}, our resolution in both velocity and overshoot is quite coarse, so we do not expect to place tight constraints on $\theta$ Oph in this exploratory calculation.  We fixed the metallicity of our models at Z = 0.02, using the \citet{GS} abundances.  For our asteroseismic comparison, we fixed the mass at 9.5 \msun, after calculating the evolutionary tracks of 8.5, 9 and 9.5 \msun models and determining which models gave the best match to the observed location of $\theta$ Oph.    

The seven observed frequencies of $\theta$ Oph were compared against the 9.5 \msun\ models shown in Figure \ref{fig:HR}, evolved to an age of 15.6 Myr.  We used the spectroscopic mode identifications given by \citet{briquet05}, but used only the given $\ell$ values as constraints.  As for the calculations performed in Section \ref{idit}, we found that the best results were obtained when each frequency is compared only to model frequencies of the same $\ell_o$.  When we do this comparison for $\theta$ Oph, we find the best fit model has a rotation and overshoot of (35 $\pm$ 35 \kms , 0.28 $\pm$ 0.1H$_p$).  This model is located within the observed photometric errors for $\theta$ Oph, with a luminosity of $log(L/L_{\odot}) = 3.75$ and an effective temperature $T_{eff} = 22590 K$.  The rotation velocity found in our best solution is in good agreement with the rotation velocity as determined from both the $vsini$ and the observed mode splitting, 29 $\pm$ 7 \kms\ \citep{briquet05,briquet07}.    The convective core overshoot determined here is slightly lower than that determined by \citet{briquet07}, although the two results do agree within the errors.  It seems that when a more accurate treatment of rotation is taken into account, at least for slowly rotating stars, this does indeed reduce the need for convective core overshooting.  Our new estimate is still higher than the results determined for other $\beta$ Cephei stars (typically 0.1-0.2), although again, the results agree to within our uncertainties.  
 
\begin{figure}
\resizebox{\hsize}{!}{\includegraphics{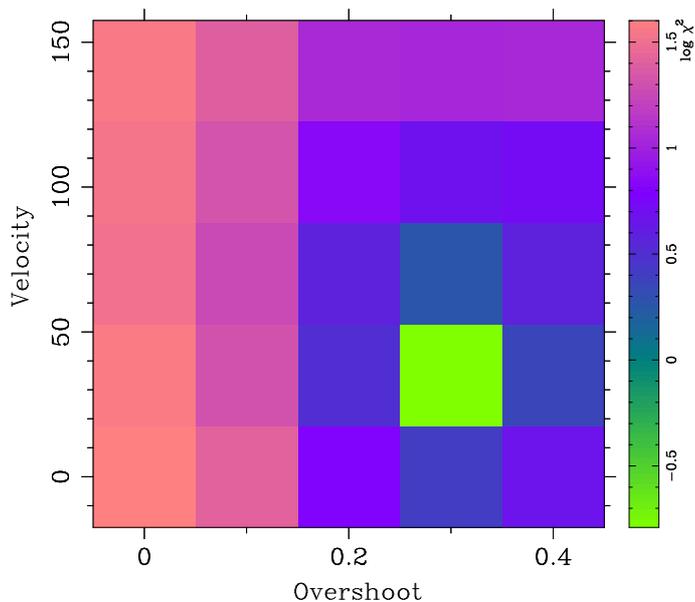}}
\caption{\label{fig:thetaoph}Log $\chi^2$ plotted as a function of velocity (y-axis) and convective overshoot (x-axis).  The minimum (green) in $\chi^2$ for this star is for $\alpha$ = 0.28, with a velocity of 35 \kms.}
\end{figure}  

Our models for $\theta$ Oph have been calculated using a metallicity of Z = 0.02 at the composition of \citet{GS}, considerably higher than that determined by \citet{briquet07}.  As discussed in their work, higher metallicity is correlated to lower convective core overshoot, so one would expect our models to have a lower overshoot.   In fact, based on the results given in their Table 4, we would expect a 1D model with our metallicity to have an overshoot of $\alpha_{ov}$ = 0.32. As is expected, this is lower than the overshoot calculated by \citet{briquet07}, and is much closer to the value obtained with our 2D models.  Nevertheless, our models also show the need for an unusually high convective core overshoot in this star.  Although our models have shown some indications that rotation and overshoot can be complimentary effects, it appears that more rapid rotation is required in order to influence the frequencies.  

Another limitation of our models is the restriction to fixed mass and age.  This is of course not realistic, and has been introduced to limit the computations required.  As mentioned in Section \ref{method}, we initially calculated models at three different masses, 8.5, 9 and 9.5 \msun.  All of these models were close to the observed error box of $\theta$ Oph, although the lower mass models needed to be evolved to a higher age.  Although detailed frequency calculations were not performed for these models, it seems likely that matching models could be found.  Including mass and age as free parameters is currently under investigation and will be included in future work.

\section{Conclusions}
\label{conclusions}


We have found that both rotation and overshoot do have an effect on stellar frequencies, although this is predominately through the effects of changing stellar radius.  Increasing convective core overshoot increases the unscaled frequencies, while rotation causes the unscaled frequencies to decrease, although certain choices of scaling can minimize this effect for rotation.  Although the frequencies themselves may change, as expected we find that the overshoot has no effect on the mode splitting.  

We also investigated the effect of overshooting on the large and small separations.  We find that increasing convective core overshoot  decreases the large separations, but increases the small separation.  Unfortunately, both of these effects are similar to the effects of rotation noted by \citet{me09}, and may not be useful for disentangling the effects of rotation and overshoot.  

We have attempted to use the changes in frequency to constrain the best fitting convective core overshoot.  Using individual modes, we were able to easily constrain the convective core overshoot, although not the rotation.  As should be expected, different modes gave different results, so we developed a modified $\chi^2$ statistic, simultaneously fitting all known frequencies for models of known mass and age (9.5 \msun, 15.6 Myr).  We found this was most effective if the $\ell_o$s of all frequencies were known.  When the $\ell_o$s are included, we are able to correctly determine both rotation and overshoot for models in our grid as well as for three test models with slightly different $v_{eq}$ and $\alpha_{ov}$.  Although the uncertainties on our results are large as a result of the coarse grid spacing used, in all cases the best solutions surrounded the true parameters of the test models.    

We also conducted a hare and hound exercise using frequencies calculated using 2D stellar models and pulsation calculations.  We found that using 1D stellar models, we were able to find a model that reproduced the frequencies for reasonably close values of age, mass, rotation and convective core overshoot.  The 1D models were able to find a velocity quite close to the true rotation rate, although the determination of core overshoot was not as accurate.  
For a model with true parameters (108, 0.25), the 1D models returned a best fit of (125, 0.1), while the 2D models found a best fit of (145, 0.28).  The mass and age returned by the 1D models were also quite close to the true values.   

Finally, we applied these methods to $\theta$ Oph, a rotating $\beta$ Cephei star with seven observed frequencies.  Using models at a fixed mass and metallicity, we find an overshoot slightly lower than that determined by \citet{briquet07}, with $\alpha$ = 0.28 $\pm$ 0.1, as expected for our higher metallicity models.  The rotation velocity for this model agrees well with the observed value, around 30 \kms.  We also find that a good match to $\theta$ Oph requires a more massive star than determined by \citet{briquet07}, around 9.5 \msun\ vs 8.2 \msun\ in their models.  This difference in mass could be a a consequence of including rotation in our models, which tends to make models appear less massive, but may also be a result of the higher metallicity used in our models.  Decreasing the convective core overshoot from 0.44 to 0.28 H$_p$ brings $\theta$ Oph closer to the range of overshoots found in other $\beta$ Cephei stars, around 0.1-0.2 H$_{p}$, but is still high.  

As discussed above, (Section \ref{HH}), the fact that the 1D calculation returns a lower overshoot than the 2D models while the 1D calculations find a higher overshoot for $\theta$ Oph is probably a result of the 1D treatment of rotation.  At slow rotation velocities ($<$ 100 \kms), like $\theta$ Oph, the 1D method works well, while for more rapidly rotating models, as in the hare and hound exercise performed here, the 1D treatment of rotation results in an underestimate of the convective core overshoot.

\begin{acknowledgements}
The authors acknowledge financial support from the French National Research Agency (ANR) for the SIROCO (SeIsmology, ROtation and COnvection with the COROT satellite ) project.
\end{acknowledgements}

\bibliographystyle{aa}
\bibliography{13855}

\end{document}